\documentclass{PoS}

\title{Quark mass determinations with the RI-SMOM scheme and HISQ action}

\ShortTitle{Quark mass determinations with the RI-SMOM scheme and HISQ action}

\author{A. T. Lytle\\
        INFN, Sezione di Roma Tor Vergata, Via della Ricerca Scientifica 1, 00133 Roma RM, Italy\\
        E-mail: \email{andrew.lytle@glasgow.ac.uk}}

\author{C. T. H. Davies\\
        SUPA, School of Physics and Astronomy, University of Glasgow, Glasgow, G12 8QQ, UK\\
        E-mail: \email{christine.davies@glasgow.ac.uk}}

\author{\speaker{D. Hatton}\\
        SUPA, School of Physics and Astronomy, University of Glasgow, Glasgow, G12 8QQ, UK\\
        E-mail: \email{d.hatton.1@research.gla.ac.uk}}

\author{G. P. Lepage\\
        Laboratory for Elementary-Particle Physics, Cornell University, Ithaca, New York 14853, USA\\
        E-mail: \email{gplepage@gmail.com}}

\author{C. Sturm \\
        Institut f{\"u}r Theoretische Physik und Astrophysik, Universit{\"a}t W{\"u}rzburg, Emil-Hilb-Weg 22, D-97074 W{\"u}rzburg, Germany\\
        E-mail: \email{christian.sturm@physik.uni-wuerzburg.de}}

\author{HPQCD Collaboration\\
        www.physics.gla.ac.uk/HPQCD}

\abstract{Lattice QCD provides several avenues for the high precision determination of quark masses. Using the RI-SMOM scheme applied to lattice calculations with the HISQ action, we obtain mass renormalisation factors that we use to provide strange and charm quark masses with 1\% precision. The calculation involves the study of various sources of systematic uncertainty, including an analysis of possible nonperturbative (condensate) contributions. These results allow a comparison of different mass determination methods of comparable precision. In particular we (HPQCD) find good agreement between RI-SMOM and current-current correlator determinations based on the same lattice QCD bare masses, providing a strong test of our understanding of systematic uncertainties.}

\FullConference{The 36th Annual International Symposium on Lattice Field Theory - LATTICE2018\\
		22-28 July, 2018\\
		Michigan State University, East Lansing, Michigan, USA.}

\begin{document}

\section{Motivation}

It is well demonstrated that lattice QCD is a powerful framework for the extraction of Standard Model (SM) parameters to high precision. Such work is important as input for various SM calculations used in high energy experiments such as the ongoing work at the LHC \cite{Lepage:2014fla}. Presented here is a summary of a calculation of the strange and charm quark masses by tuning of lattice computations to experimentally measured meson masses with renormalisation of the resulting bare quark masses in the RI-SMOM scheme, with a final perturbative matching to $\overline{\mathrm{MS}}$, detailed in \cite{Lytle:2018evc}. This method is able to produce percent level precision, comparable to that achieved by the different methodologies of \cite{Chakraborty:2014aca} and \cite{Bazavov:2018omf}. In particular the work of \cite{Chakraborty:2014aca} uses much of the same input data (bare quark masses and configurations) and this separate determination therefore provides a strong check on the understanding of the (different) systematics in both calculations.

\section{The RI-SMOM scheme}

The RI-SMOM scheme \cite{Sturm:2009kb} defines renormalisation factors in terms of (Landau) gauge fixed inverse propagators and vertex functions in momentum space, on which renormalisation conditions are imposed. For example, the wavefunction renormalisation $Z_q$ is fixed to be 1 in the free theory. All vertex renormalisation factors are calculated as the ratio of $Z_q$ and the relevant amputated vertex function $\Lambda_{\mathcal{O}}$, with $\mathcal{O}$ denoting the operator at the vertex. These vertex functions have a symmetric kinematic setup with $q^2=p_1^2=p_2^2 \equiv \mu^2$ where $p_1$ and $p_2$ are the ingoing and outgoing quark momenta and $q=p_1-p_2$ is the momentum insertion at the vertex.

This scheme can be implemented nonperturbatively on the lattice as long as care is taken in the consideration of condensate contributions which are not present in the perturbative matching calculations to other schemes, performed in the continuum. For the purposes presented here this is done (as is typical) at vanishing valence quark mass which is achieved through an extrapolation to this point using multiple valence masses (see Section \ref{sec-mass-extrap} and Section IV B of \cite{Lytle:2018evc}).

Here the mass renormalisation factors $Z_m$ are calculated on 2+1+1 HISQ \cite{Follana:2006rc} configurations generated by the MILC collaboration \cite{Bazavov:2010ru,Bazavov:2012xda} from the scalar vertex function and propagator, using lattice spacings in the range $\sim 0.06-0.12$ fm. $Z_m$ is simply the inverse of $Z_S$. Calculations were done at multiple values of $\mu$ ranging from 2 to 5 GeV.

The RI-SMOM scheme is detailed in \cite{Sturm:2009kb} and the framework for the implementation of such schemes with staggered quarks was developed in \cite{Lytle:2013qoa}.

\section{Systematic checks}

\begin{figure}
  \centering
  \includegraphics[width=0.6\textwidth]{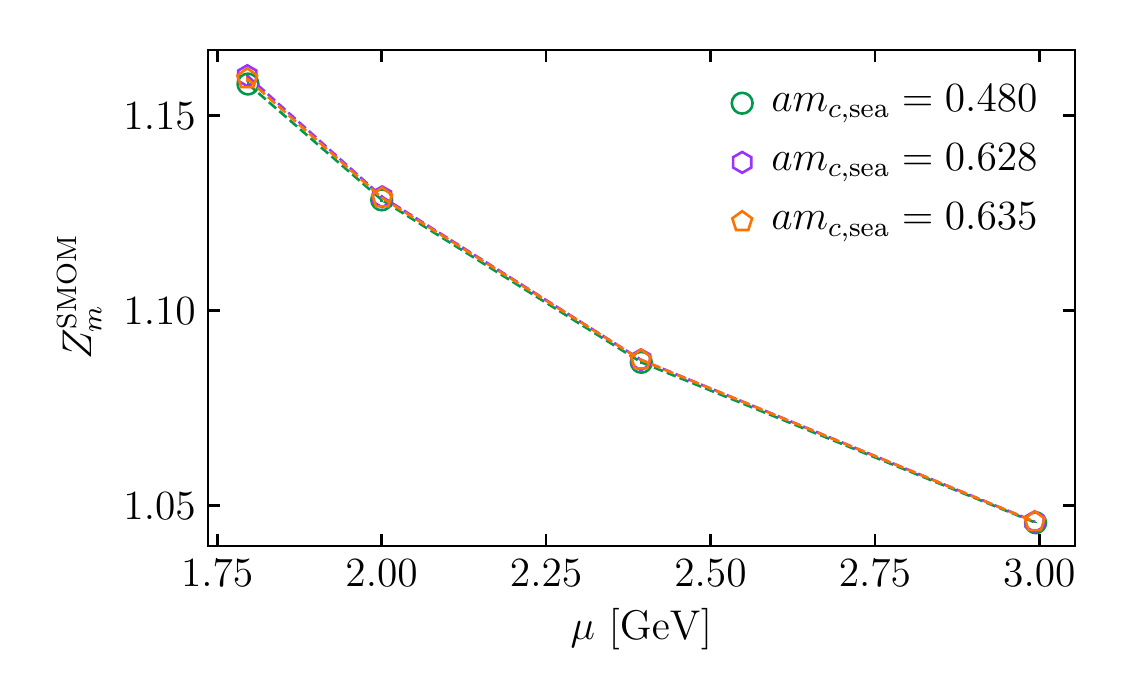}
  \caption{Dependence of $Z_m^{\mathrm{SMOM}}$ on the charm sea mass as a function of $\mu$. No significant variation can be seen.}
  \label{sea-mass}
\end{figure}

\begin{figure}
  \centering
  \includegraphics[width=0.6\textwidth]{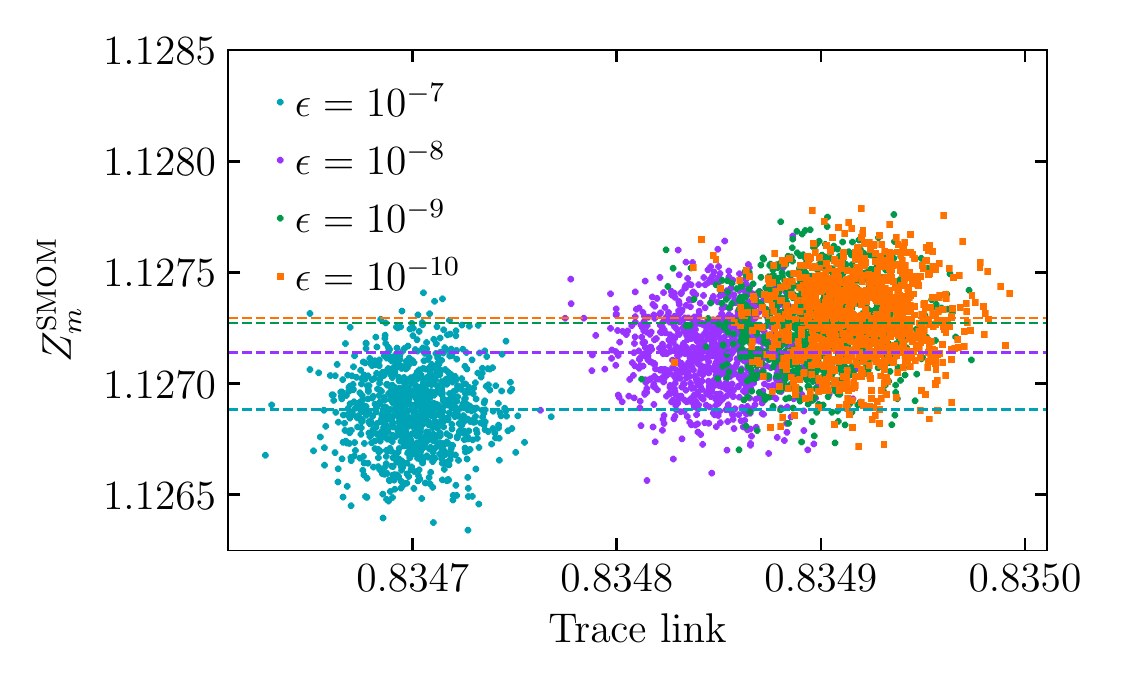}
  \caption{Scatter plot of bootstrap samples of $Z_m^{\mathrm{SMOM}}$ against the average trace link variable for different Landau gauge fixing tolerances. The mean values are shown by the dashed horizontal lines and can be seen to move substantially from tolerance $10^{-7}$ to $10^{-10}$.}
  \label{gfixing}
\end{figure}

In order to assess the level at which finite volume and mistuned or unphysical sea quark masses affect the results, $Z_m$ was calculated on lattice ensembles with multiple volumes and multiple sea quark masses. No significant variation was seen between ensembles, an example being given by Figure \ref{sea-mass}. This is as expected from ultraviolet quantities. However, a significant (although small) and momentum dependent effect was observed as the tolerance of Landau gauge fixing was varied. The various data used in this analysis used a gauge fixing tolerance of either $10^{-7}$ or $10^{-14}$ with a $\mu$ dependent systematic error being added to the former to account for the effect seen in Figure \ref{gfixing} which shows the spread of bootstrap data samples as the tolerance is tightened.

As the SMOM to $\overline{\mathrm{MS}}$ matching calculations have been done at zero mass they neglect the effect of sea quark masses which are present in lattice calculations. We use matching factors that account for the massive charm in the sea for which details are provided in Appendix A of \cite{Lytle:2018evc}.

\section{Mass extrapolation} \label{sec-mass-extrap}

\begin{figure}
  \centering
  \includegraphics[width=0.6\textwidth]{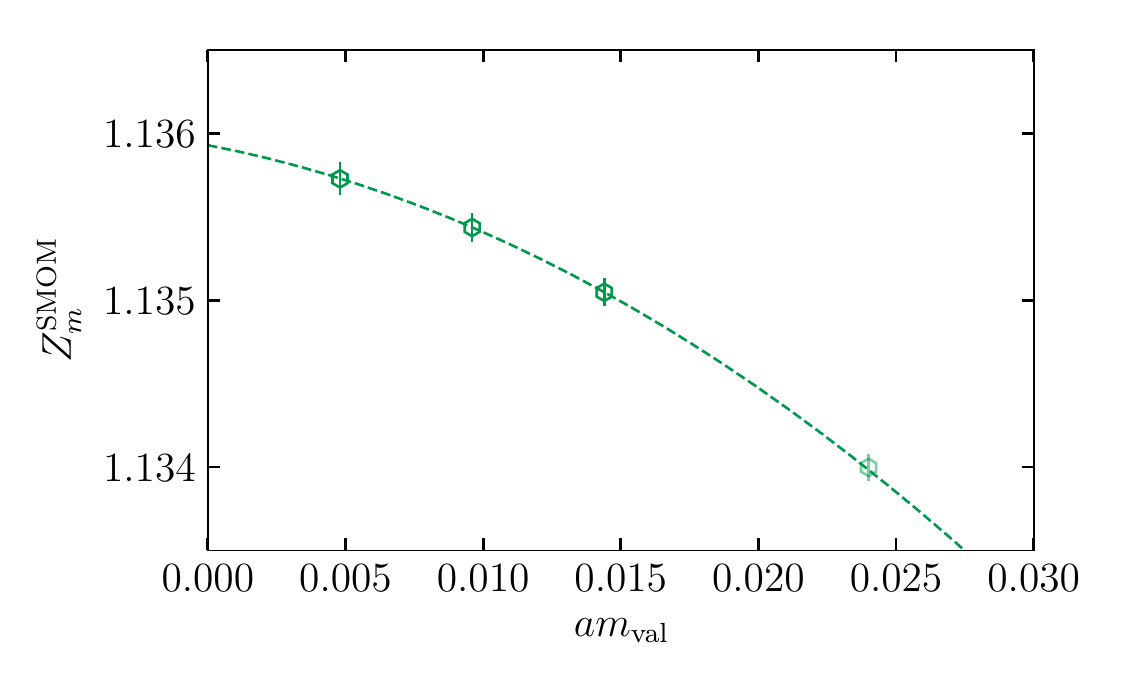}
  \caption{Valence mass extrapolation at $\mu = 3\ \mathrm{GeV}$ on a lattice with a spacing of approximately 0.06 fm. The highest mass point is not included in the fit.}
  \label{mass-extrap}
\end{figure}

In order to extrapolate to the zero valence mass point the calculation of $Z_m$ was performed at three different valence masses and then extrapolated using a polynomial in $am_{\mathrm{val}}$ up to third order. This was found to give good $\chi^2$ for all fits and can be demonstrated to provide an accurate representation of the mass dependence even up to the strange quark mass, as displayed in Figure \ref{mass-extrap} where the highest mass point was not included in the fit shown. All data used in final fits has been extrapolated to zero valence quark mass. We have been careful to propagate the effect of correlations between different $\mu$ values on a given ensemble through the calculation.

\section{Condensate contributions}

The extrapolation of the valence mass removes the contributions of chiral condensates that appear in the operator product expansion proportional to the mass. However, there will still be contributions from gluon condensates. The leading gauge invariant such condensate $\langle G_{\mu\nu}G^{\mu\nu} \rangle$ is expected to be small and will be suppressed by $\mu^4$ but the gauge noninvariant quantites that we use here get contributions from gauge noninvariant condensates, most prominently $\langle A^2 \rangle$ which will only be suppressed by $\mu^2$ \cite{Chetyrkin:2009kh}. Such a condensate is therefore allowed for in the final continuum extrapolation fit. This has been neglected in previous calculations of $Z_m$ using this method.

\section{Continuum extrapolation and extraction of quark masses}

The data used in the final fit is constructed by multiplying the bare quark mass on each lattice by the relevant $Z_m^{\mathrm{SMOM}}$, then multiplying by the $\overline{\mathrm{MS}}$ matching factor and running to a reference scale of 3 GeV in the $\overline{\mathrm{MS}}$ scheme. Having all data points perturbatively run to the same reference scale allows the effects of condensates to become visible.

The continuum extrapolation employed here accounts for sources of discretisation errors in both the bare masses and $Z_m$ as well as condensate contributions, residual sea quark mass dependence and neglected $\alpha_s^3$ terms in the matching to $\overline{\mathrm{MS}}$. These extrapolations (done separately for the strange and charm masses) result in determinations with 1\% precision whose values are in good agreement with the determinations of \cite{Chakraborty:2014aca} and \cite{Bazavov:2018omf}.

\begin{figure}
  \centering
  \includegraphics[width=0.6\textwidth]{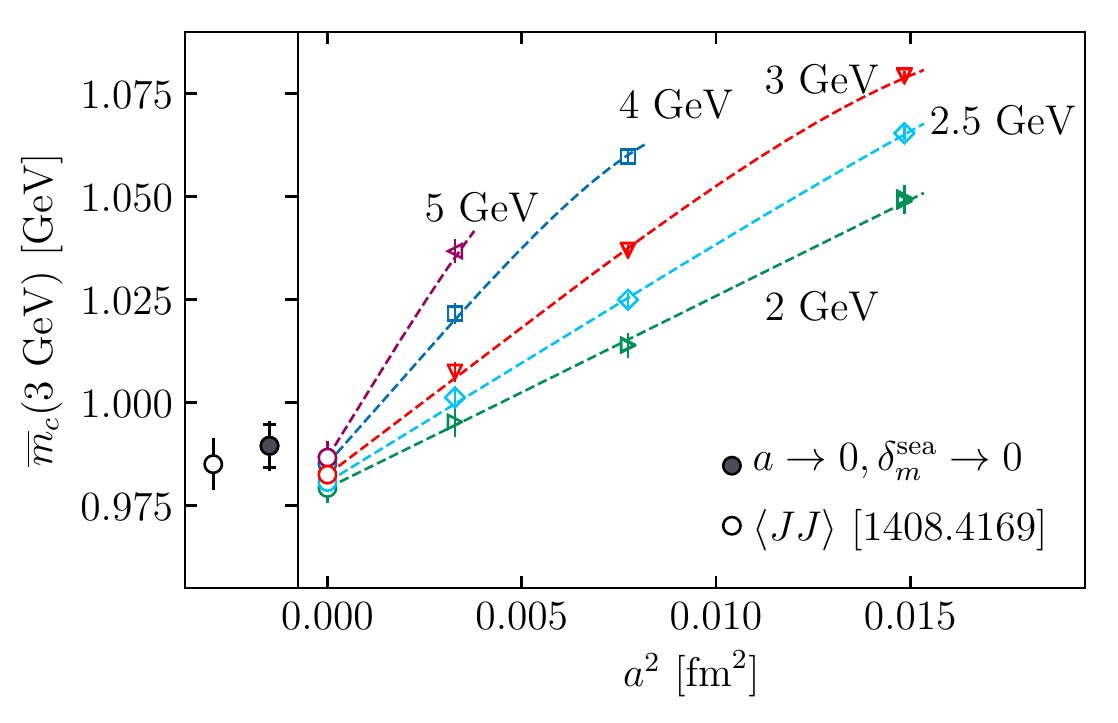}
  \includegraphics[width=0.6\textwidth]{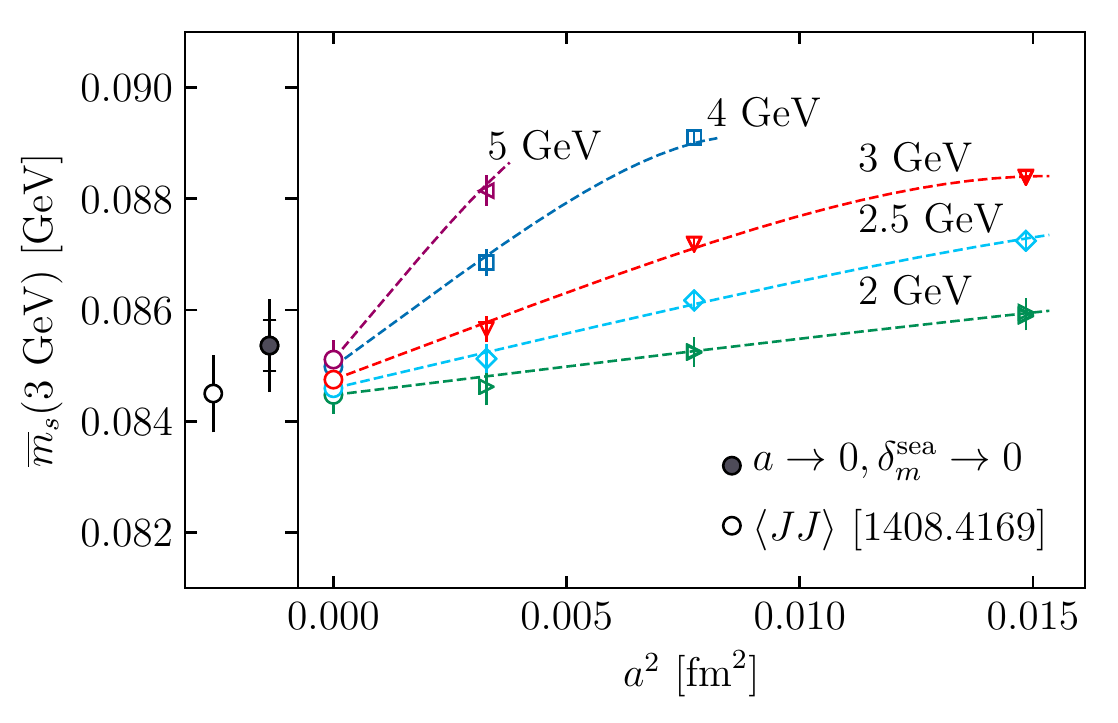}
  \caption{Continuum extrapolations of $\overline{m}_c$ and $\overline{m}_s$ with a fit accounting for condensates and sea mass mistunings. The circles at the end of the fit lines have the sea mass mistunings removed but still contain condensate contributions, while the filled circle on the left is the final answer, with condensate contributions removed. The empty circle on the far left is the result from \cite{Chakraborty:2014aca}.}
  \label{cont-extraps}
\end{figure}

The greatest source of uncertainty in the final values (particularly for the strange quark) are the correlated and uncorrelated uncertainties on the bare quark masses. The uncorrelated uncertainties arise from the independent tuning done on each ensemble while the correlated uncertainities come from fitting and the $w_0$ value \cite{Dowdall:2013rya} used in each lattice spacing determination. After that the next largest sources are from the continuum extrapolation and from the condensate terms. Both could in principle be reduced by using finer lattices which allow a move to higher values of $\mu$, which will lead to further suppression of the condensate contributions.

\section{Results}

The final results in the $\overline{\mathrm{MS}}$ scheme at a reference value of 3 GeV are

\begin{eqnarray}
  \overline{m}_c(3\ \mathrm{GeV},n_f=4) = 0.9896(61)\ \mathrm{GeV} \\
  \overline{m}_s(3\ \mathrm{GeV},n_f=4) = 0.08536(85)\ \mathrm{GeV} . \nonumber
\end{eqnarray}

These can be run to the conventionally quoted scales of $\overline{m}_c$ and 2 GeV respectively, giving

\begin{eqnarray}
  \overline{m}_c(\overline{m}_c,n_f=4) = 1.2757(84)\ \mathrm{GeV} \\
  \overline{m}_s(2\ \mathrm{GeV},n_f=4) = 0.09449(96)\ \mathrm{GeV} . \nonumber
\end{eqnarray}

These results can be combined as described in \cite{Lytle:2018evc} with other available $n_f=4$ lattice determinations to give world averages of
\begin{eqnarray}
  \overline{m}_c(\overline{m}_c,n_f=4)_{2+1+1\ \mathrm{av.}} = 1.2753(65)\ \mathrm{GeV} \\
  \overline{m}_s(2\ \mathrm{GeV},n_f=4)_{2+1+1\ \mathrm{av.}} = 0.09291(78)\ \mathrm{GeV} . \nonumber
\end{eqnarray}

In summary, good 1\% level precision agreement for the strange and charm quark masses has been achieved for different methodologies with different sources of systematic uncertainty, indicating good control over these uncertainties for lattice QCD quark mass determinations.

\textbf{Acknowledgments} We are grateful to MILC for the use of their gluon field ensembles. This
work was supported by the UK Science and Technology Facilities Council. The calculations used
the DiRAC Data Analytic system at the University of Cambridge, operated by the University of
Cambridge High Performance Computing Service on behalf of the STFC DiRAC HPC Facility
(www.dirac.ac.uk). This is funded by BIS National e-infrastructure and STFC capital grants and
STFC DiRAC operations grants.


\begin{thebibliography}{99}

  \bibitem{Lepage:2014fla}
    G.~P.~Lepage, P.~B.~Mackenzie and M.~E.~Peskin,
    arXiv:1404.0319 [hep-ph].

  \bibitem{Lytle:2018evc}
    A.~T.~Lytle {\it et al.} [HPQCD Collaboration],
    Phys.\ Rev.\ D {\bf 98} (2018) no.1,  014513
    doi:10.1103/PhysRevD.98.014513
    [arXiv:1805.06225 [hep-lat]].

  \bibitem{Chakraborty:2014aca}
    B.~Chakraborty {\it et al.},
    Phys.\ Rev.\ D {\bf 91} (2015) no.5,  054508
    doi:10.1103/PhysRevD.91.054508
    [arXiv:1408.4169 [hep-lat]].

  \bibitem{Bazavov:2018omf}
    A.~Bazavov {\it et al.} [Fermilab Lattice and MILC and TUMQCD Collaborations],
    Phys.\ Rev.\ D {\bf 98} (2018) no.5,  054517
    doi:10.1103/PhysRevD.98.054517
    [arXiv:1802.04248 [hep-lat]].

  \bibitem{Sturm:2009kb}
    C.~Sturm, Y.~Aoki, N.~H.~Christ, T.~Izubuchi, C.~T.~C.~Sachrajda and A.~Soni,
    Phys.\ Rev.\ D {\bf 80} (2009) 014501
    doi:10.1103/PhysRevD.80.014501
    [arXiv:0901.2599 [hep-ph]].

  \bibitem{Follana:2006rc}
    E.~Follana {\it et al.} [HPQCD and UKQCD Collaborations],
    Phys.\ Rev.\ D {\bf 75} (2007) 054502
    doi:10.1103/PhysRevD.75.054502
    [hep-lat/0610092].

  \bibitem{Bazavov:2010ru}
    A.~Bazavov {\it et al.} [MILC Collaboration],
    Phys.\ Rev.\ D {\bf 82} (2010) 074501
    doi:10.1103/PhysRevD.82.074501
    [arXiv:1004.0342 [hep-lat]].

  \bibitem{Bazavov:2012xda}
    A.~Bazavov {\it et al.} [MILC Collaboration],
    Phys.\ Rev.\ D {\bf 87} (2013) no.5,  054505
    doi:10.1103/PhysRevD.87.054505
    [arXiv:1212.4768 [hep-lat]].

  \bibitem{Lytle:2013qoa}
    A.~T.~Lytle and S.~R.~Sharpe,
    Phys.\ Rev.\ D {\bf 88} (2013) no.5,  054506
    doi:10.1103/PhysRevD.88.054506
    [arXiv:1306.3881 [hep-lat]].

  \bibitem{Chetyrkin:2009kh}
    K.~G.~Chetyrkin and A.~Maier,
    JHEP {\bf 1001} (2010) 092
    doi:10.1007/JHEP01(2010)092
    [arXiv:0911.0594 [hep-ph]].

  \bibitem{Dowdall:2013rya}
    R.~J.~Dowdall, C.~T.~H.~Davies, G.~P.~Lepage and C.~McNeile,
    Phys.\ Rev.\ D {\bf 88} (2013) 074504
    doi:10.1103/PhysRevD.88.074504
    [arXiv:1303.1670 [hep-lat]].


\end{thebibliography}
\end{document}